\documentclass[twocolumn,amsmath,aps,fleqn]{revtex4}

\usepackage{amssymb}
\usepackage{amsmath}
\usepackage{epsfig}
\usepackage{subfigure}
\usepackage{mathrsfs}
\usepackage{longtable}

\newcommand{\be}{\begin{eqnarray}}
\newcommand{\ee}{\end{eqnarray}}
\newcommand{\bea}{\begin{eqnarray}}
\newcommand{\eea}{\end{eqnarray}}

\newcommand{\grad}{\ensuremath{\vec{\nabla}}}
\newcommand{\adotoa}{\ensuremath{{\cal H}}}
\newcommand{\Uc}{\ensuremath{{\cal U}}}
\newcommand{\Vc}{\ensuremath{{\cal V}}}

\begin{document}

\title{The linear growth rate of structure in Parametrized Post Friedmannian Universes}
\author{Pedro G. Ferreira}
\email{pgf@astro.ox.ac.uk}
\affiliation{Oxford Astrophysics, Physics, DWB, Keble Road, Oxford, OX1 3RH, UK}
\author{Constantinos Skordis}
\email{skordis@nottingham.ac.uk}
\affiliation{School of Physics and Astronomy, University of Nottingham, University Park, Nottingham,  NG7  2RD,UK}


\begin{abstract}
A possible solution to the dark energy problem is that Einstein's theory of general relativity is modified. A suite of models have been proposed that, in general, are unable to predict the
correct amount of large scale structure in the distribution of galaxies or anisotropies in the Cosmic Microwave Background. It has been argued, however, that it should be possible to constrain
a {\it general} class of theories  of modified gravity by focusing on properties such as the growing mode, gravitational slip and the effective, time varying Newton's constant. We show that
assuming certain physical requirements such as stability, metricity and gauge invariance, it is possible to come up with consistency conditions between these various parameters. In this paper we focus on theories which have, at most, 2$^{\rm nd}$ derivatives in the metric variables and 
find restrictions that shed light on current and future
experimental constraints without having to resort to a (as yet unknown) complete theory of
modified gravity. We claim that future  measurements of the growth of structure on
small scales (i.e. from 1-200 $h^{-1}$ Mpc) may lead to tight constraints on both dark energy and modified theories of gravity.
\end{abstract}

\maketitle


\section{Introduction}
The {\it dark energy} problem, i.e. the possibility that 70$\%$ of the Universe seems to be permeated by an invisible fluid which behaves repulsively under gravity and 
does not cluster, has been the focus of research in cosmology for over decade. There are
a host of proposals \cite{CopelandSameTsujikawa2006} and a battery of experiments are under way,
or on the drawing board, to characterize the nature of this elusive source of energy \cite{SKA,RefregierEtAl2010,DES,LSST}.

In recent years, an alternative possibility has emerged, that Einstein's General theory of relativity is incorrect on cosmological scales and must be modified. Although the idea that General Relativity
is incomplete has been around since the early 1960s \cite{Dirac1938a,Jordan1949,BransDicke1961,Sakharov1968a}, there are now a number of proposals for what this theory of modified gravity might be \cite{FerreiraStarkman2009}. 
The Einstein-Hilbert action, $S_g\propto\int d^4x \sqrt{-g}R$ (where $g$ is the metric determinant and $R$ is the scalar curvature of 
a metric $g_{ab}$) can be replaced by
a more general form $S_g\propto\int d^4x \sqrt{-g}F(R)$ where $F$ is an appropriately chosen function of $R$ \cite{Amendola07a,SotiriouFaraoni2010a}; the
 dynamics of the gravitational field can emerge from a theory in higher dimensions such as one might encounter in 
brane worlds \cite{DvaliGabadadzePorrati2000d}; a preferred reference frame may emerge from the spontaneous symmetry breaking of 
local Lorentz 
symmetry \cite{ZlosnikFerreiraStarkman2007,JimenezMaroto2008,DimopoulosKarciauskasWagstaff2010a,KoivistoMota2008}; the metric which satisfies 
the Einstein equation is not necessarily the one that defines  geodesic motion \cite{Bekenstein1993} 
but is related to a second metric via additional fields~\cite{Bekenstein2004a,Skordis2008a,Skordis2009a}
or connections~\cite{Banados2008,BanadosEtAl2008,Milgrom2009}; the Einstein-Hilbert action may be deformed by choosing as fundamental variables of gravity, $SU(2)$ connections~\cite{Plebanski1977,Krasnov2008,KrasnovShtanov2010}.

Many of these models have been successful in reproducing, for example, the observed relation between redshift and luminosity distances from distant supernovae.
 They have, however, generally failed to reproduce the observed clustering of galaxies on large scales as well as the 
anisotropies in the Cosmic Microwave Background (CMB) unless the modified theory becomes
effectively equivalent to general relativity (i.e. the Einstein-Hilbert action and a cosmological constant), e.g.~\cite{BanadosFerreiraSkordis2008,ZuntzFerreiraZlosnik2008,ZuntzEtAl2010}. 
The general problem that seems to plague most theories is an excess of
power on the very largest scales which manifests itself through the Integrated Sachs Wolfe
(ISW) effect and a mismatch between the normalization of the power spectrum of fluctuations on
the largest and smallest scales. As yet, a truly compelling and viable model of modified theory
of gravity has yet to be but forward which may resolve the dark energy problem.

All is not lost, however, and progress can be made in learning about potential modifications
to gravity by extracting phenomenogical properties that can be compared to observation- the "Parametrized Post Friedmannian" approach \cite{HuSawicki2007a}.  In this paper we 
focus on a key observable characterizing the evolution of large scale structure:
the growing mode of gravitational collapse.  

\begin{figure}[htbp]
\begin{flushleft}
\vspace{-15pt}
\epsfig{figure=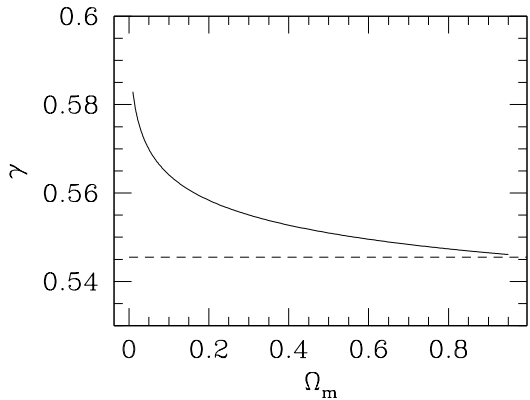,width=9cm}
\end{flushleft}
\vspace{-100pt}
\caption{The solid line is the growth parameter, $\gamma$, for a $\Lambda$CDM universe, as
a function of $\Omega_M$. For small values of $\Omega_\Lambda$, $\gamma$ is well
approximated by $6/11$ (dashed line) but there are deviations as $\Omega_\Lambda$ grows; we find errors of $0.7\%$, $3.3\%$ and $4.2\%$ when $\Omega_M=0.7$, $0.3$ and $0.05$.}
\label{lambda}
\vspace{-10pt} 
\end{figure}

The time evolution of the density field can be
a sensitive probe of not only the expansion rate of the the Universe but also its matter content.
In a flat, matter dominated universe we have that $\delta_M$, the density contrast of matter, evolves as $\delta_M\propto a$ where  $a$ is the scale factor of the Universe. We can parametrize
deviations from this behaviour in terms of $\gamma$  \cite{Peebles1980,Linder2005a,Lee09a,Lee09b} through
\begin{eqnarray}
\gamma\equiv\frac{\ln\left[\frac{{\dot \delta}_M}{ \adotoa \delta_M}\right]  }{\ln \Omega_M}
\end{eqnarray} 
where $\Omega_M$ is the fractional density of matter, 
$\adotoa=\frac{\dot a}{a}$ 
and overdots are derivatives with regards to conformal time, $\tau$.
 For standard growth in the presence of a cosmological constant, one has that  $\gamma\approx6/11$ to a very good approximation.
This is not true over a wide range of values for $\Omega_M$. In fact, in Figure \ref{lambda}
we can see that $\gamma$ deviates from its early-universe asymptotic value as $\Omega_M\rightarrow 0$. A natural question to ask is how $\gamma$ depends on different aspects of the Universe
and how might use it to constrain dark energy and modifications to gravity. In this
paper we will focus on a few of these properties.

One important
property of the Universe is the {\it equation of state} of dark energy, characterized by the constant
(or function of time), $w$:
\begin{eqnarray}
P_E=w\rho_E.
\label{fluid_eq_state}
\end{eqnarray}
$P_E$ and $\rho_E$ are the pressure and energy densities of the dark energy.
The function $w$ may be time varying and is related to the adiabatic speed of sound $c_a^2$ as
\begin{equation}
c_a^2 = w - \frac{\dot{w}}{3\adotoa(1+w)} 
\end{equation}

Another important property is {\it gravitational slip}, $\zeta$, which is normally defined to be
\begin{eqnarray}
\Phi-\Psi\equiv\zeta\Phi
\end{eqnarray}
where we are taking a linearly perturbed metric in the conformal
Newtonian gauge,
\begin{equation}
 ds^2=-a^2(1+2\Psi)d\tau^2+a^2(1-2\Phi)d\vec{x}^2.
\end{equation}
Such a parametrization has been advocated in a number of papers on modified gravity
\cite{Bertschinger2006,BertschingerZukin2008,DanielEtAl2009,DanielEtAl2010,PogosianEtAl2010,Bean2009,ReyesEtAl2010} and it has been shown that it can lead to a number of 
observational effects. Albeit simple, and appealing, such a parametrization of slip is not necessarily general and, as we shall see in the next section, necessarily implies other non-trivial
modifications to the gravitational sector. Such modifications are, in general, not explicitely
acknowledged but may correspond to unexpected assumptions about 
any putative, underlying theory. Hence a more general assumption (at least within the context of 2$^{\rm nd}$ order
theories) would be that gravitational slip would depend on $\Phi$ and ${\dot \Phi}$  (this is explained in more detail in section-\ref{sec_consistency}
and in ~\cite{Skordis2008b})

Finally, we can define an {\it effective Newton's constant} in the relativistic Newton-Poisson
equation
\begin{eqnarray}
\nabla^2\Phi=4\pi a^2G_{eff}\sum_X\rho_X[\delta_X+3(1+w_X)\frac{\dot a}{a}\theta_X] \nonumber
\label{eq_eff_Poisson}
\end{eqnarray}
where $\delta_X$ is the density contrast and $\theta_X$ is the momentum of the cosmological fluid $X$ which has an equation of state $w_X$. We can define the dimensionless function
\begin{eqnarray}
\mu^2\equiv\frac{G}{G_{eff}}
\end{eqnarray}
where $G$ is the "bare" Newton constant.
 It then makes sense to
try to constrain ($\gamma$, $w$, $\zeta$ and $\mu$) in the hope that it may be possible to shed light on a possible theory of modified gravity.

Although there alternative proposals \cite{KoivistoMota2006,Mota2007}, a number of groups have have pioneered the use of this simple parametrization of modified gravity
(in terms of $\zeta$, $\mu^2$ or both): in \cite{Bertschinger2006,BertschingerZukin2008} it was argued that
gravitational slip might be a generic prediction for modified theories of gravity, in \cite{DanielEtAl2009,DanielEtAl2010} it
was shown that it would be possible to constrain it through cross correlations of  the CMB with galaxy surveys and in \cite{Hu2008} from the ISW effect; 
weak lensing has been proposed as a possible route for constraining these parameters \cite{UzanBernardeau2001,ZhangEtAl2007,Schmidt2009a} with a tentative detection of modification being proposed in \cite{Bean2009,BeanTangmatitham2010}. Much is expected from applying these methods to future ambitious experiments that will map out the large scale structure of the Universe. Indeed constraints of General Relativity are a core element of the science that could be extracted from
the Euclid experiment~\cite{RefregierEtAl2010}.

Given that such an approach is phenomenological, the general attitude has been to leave these parameters completely free. There is merit to such an approach in that one isn't restricting oneself
to a particular theory and hence constraints will be general. It is true however that is is possible to
idenitfy (reasonably general) consistency conditions for  ($\gamma$, $w$, $\zeta$ and $\mu$), contingent on specific physical assumptions. In this paper we state these assumptions and
present restrictions on ($\gamma$, $w$,$\zeta$ and $\mu$). We shall use the formalism first
proposed  by one of us\cite{Skordis2008b} which spells out how to build consistent modifications to gravity. 

This paper is structured as follows. In Section \ref{formalism} we recap the formalism
presented in \cite{Skordis2008b} and relate it to the parameters we wish to study phenomenologically.
We discuss how the consistency conditions reduce the freedom to choose arbitrary ($\gamma$, $\zeta$, $\mu^2$). In Section \ref{small} we implement the consistency conditions and
find a relationship between the parameters by looking at the evolution equation for the density
contrast in matter for small wavelengths. In doing so, we find find analytic expressions for the
relationships and briefly assess the range of scale to which they are applicable. In Section \ref{General} we find analytic expressions for $\gamma$ to 2$^{\rm nd}$ order for a general
parametrization which is consistent with the Parametrized Post Newtonian approximation
on small scales.In Section \ref{Discussion} we discuss the generality of the results and how
they may be extended to other, more exotic models. 
\section{The formalism}

\label{formalism}

We now summarize the formalism, the details of which can be found
in the \cite{Skordis2008b}: we present the field equations, the evolution equations for
the fluid components and the consistency conditions for modifications to the
field equations. We shall further assume a spatially flat universe, but our results can be easily generalized to
include curvature.

\subsection{The background cosmology}
As discussed in~\cite{Skordis2008b}, the background equations for any theory of gravity for which the metric is Friedman-Robertson-Walker (FRW) 
can be recast in the usual form used in GR. The Friedman equation simply reads
\begin{equation}
3 \adotoa^2 = 8\pi G a^2 \sum_X \rho_X
\end{equation}
In addition to the Friedman equation we also have the Raychaudhuri equation $-2\frac{\ddot{a}}{a} + \adotoa^2 = 8\pi G a^2 \sum_X P_X$. With the help of the
Friedman equation, in a universe containing only pressureless matter and dark energy (as is approximately the case in the late universe) 
the  Raychaudhuri equation  may be rewritten as
\begin{equation}
\dot{\adotoa } =  -\frac{1}{2} \adotoa^2 (1 + 3w  \Omega_E)
\label{Ray}
\end{equation}

The dark energy density $\rho_E$ and $w$, may be in general a function of additional degrees of freedom, the scale factor $a$ or $\adotoa$.
For example, for $F(R)$ one gets $\rho_E = \frac{1}{2}(R F_R - F) - \frac{3\adotoa}{a^2} \dot{F}_R - \frac{3 \adotoa^2}{a^2} F_R$. 
But this explicit dependence of $\rho_E$ (or of $w$) is irrelevant. One may always
treat $\rho_E$ as a standard fluid  with a time-varying equation of state $w$ subject to energy conservation 
$\dot{\rho}_E + 3 \adotoa (1+w) \rho_E = 0$
(but note that there may be additional field equations that determine the time dependence of $w$).  In a universe containing only pressureless matter and dark energy, the energy conservation equation for dark energy can be rewritten as
\begin{equation}
\dot{\Omega}_E = -3\adotoa w \Omega_M \Omega_E
\end{equation}

Our discussion above has one important consequence: that one
 cannot distinguish modifications of gravity from ordinary fluid dark energy using observables based on FRW alone. As discussed in~\cite{Skordis2008b},
and further below, the situation changes drastically once we consider linear fluctuations.

\subsection{The field equations.}
The idea is to parameterize deviations from Einstein gravity at a linear level. Schematically
we can write the modified Einstein equations in the form
\begin{eqnarray}
{\delta G}^{mod}_{\mu\nu}={\delta G}_{\mu\nu} - \delta U_{\mu\nu}=8\pi G \delta T_{\mu\nu}+8\pi G \delta T^{DE}_{\mu\nu}
\end{eqnarray}
Note that for a tensor $F$ we use $\delta F$ to indicate a linear perturbation of $F$ and 
we assume that $\delta U_{ab}$ is made of the scalar metric perturbations and their derivatives.
Let us also stress that the background tensor corresponding to $U_{ab}$, i.e. $\bar{U}_{ab}$ vanishes. 
 We assume that "normal matter" (i.e. baryons, dark matter, neutrinos and photons) are contained in $T_{\mu\nu}$ and that dark energy, or any non metric degrees of freedom that
behave like dark energy (such as a scalar field- quintessence- or a dark fluid), are contained in $T^{DE}_{\mu\nu}$.  In this paper
we will restrict ourselves to two fluids: $T_{\mu\nu}$ is the energy-momentum tensor
for a pressureless fluid with density $\rho_M$, density contrast $\delta_M$ and momentum
$\theta_M$ while $T^{DE}_{\mu\nu}$ is the energy momentum tensor of a fluid with density $\rho_{DE}$, density contrast
$\delta_{DE}$ and $\theta_{DE}$ which can be characterized
defined in terms of (possibly time varying) equation of state and sound speed (we shall
use the approach of \cite{Hu1998} to model a quintessence like fluid with a constant
equation of state.

The field equations can be rewritten in the following form
\begin{eqnarray}
-2k^2\Phi  &=&  8 \pi G a^2 \sum_X \rho_X \left[\delta_X+3(1+w_X) \adotoa \theta_X\right]
\nonumber 
\\  
&& 
\ \ \ \ + U_{\Delta}+ 3\adotoa U_{\theta} \label{NP1}
\\
2({\dot \Phi}+ \adotoa \Psi) &=& 8\pi G a^2 \sum_X (\rho_X +P_X)\theta_X + U_{\theta}  
\label{momentum} 
\\
\frac{d}{d\tau}({\dot \Phi}+ \adotoa \Psi) &=& 4\pi Ga^2 \rho_E \Pi_E
+\frac{1}{6} U_P
+ \frac{1}{3} \nabla^2 U_\Sigma
\nonumber 
\\ 
&&
- 2 \adotoa ({\dot \Phi}+ \adotoa \Psi) 
+ (\adotoa^2 - \dot{\adotoa} )\Psi
\nonumber
\\
\Phi-\Psi&=& U_{\Sigma} \label{field}
\end{eqnarray}
As advertised, the $U$ terms contain modifications to gravity and we have used the notation from \cite{Skordis2008b}: $U_{\Delta}\equiv-a^2U^0_{\phantom{0}0}$,
$\grad_iU_\theta=-a^2U^0_{\phantom{0}i}$, $U_P=a^2U^i_{\phantom{i}i}$ and 
$[\grad^i\grad_j - \frac{1}{3} \grad^2 \delta^i_{\phantom{i}j} ] U_{\Sigma}=a^2(U^i_{\phantom{i}j}-\frac{1}{3}U^k_{\phantom{k}k}\delta^i_{\phantom{i}j})$.
Further, we have parameterized the dark energy pressure perturbation $\Pi_E \equiv \delta P_E/\rho_E$ as
\begin{eqnarray}
\Pi_E = c_{s}^2 \delta_E + 3 (c_{s}^2 - c_a^2) (1+w) \adotoa \theta_E
\end{eqnarray}
For adiabatic fluids, as is the case of radiation and CDM, $c_s = c_a$ and we get $\Pi_E = c_s^2 \delta_E$. In general, however,
$c_s\ne c_a$ and may in fact be a function of space as well as time.

\subsection{The fluid equations.}

It is convenient to define $\Delta_X=\delta_X-3(1+w_X)\Phi$ for $X=M,DE$. We then have that the equations of motion for the fluids \cite{Hu1998} are:
\begin{eqnarray}
{\dot \Delta}_M&=&-k^2\theta_M
\nonumber 
\\
{\dot \theta}_M&=&- \adotoa\theta_M+\Psi 
 \nonumber 
\\
{\dot \Delta}_E&=&
3\adotoa (w- c_s^2) \Delta_E
-(1+w) k^2 \theta_E 
 \nonumber 
\\
& &
-9(1+w)\adotoa(c_s^2-c_a^2) \left[\Phi + \adotoa \theta_E \right]
\nonumber 
\\
{\dot \theta}_E&=& (3c^2_{s}-1) \adotoa\theta_E +c^2_{s}\left(\frac{1}{1+w} \Delta_E+3\Phi \right) 
+\Psi  \label{fluid}
\end{eqnarray}

\subsection{The consistency conditions.}

\label{sec_consistency}
In principle, one should be able to choose arbitrary combinations of metric 
perturbations to go in the tensor $U$. Yet in \cite{Skordis2008b} it was argued that by assuming
a set of general properties, it is possible to restrict the form of $U$. For the purpose
of this paper,  we shall choose the general theory of gravity to satisfy the following restrictions:
\begin{enumerate}
\item {\it The fundamental geometric degree of freedom is the metric.} This encompasses
 most modified theories of gravity including the first order Palatini (torsion-less) formulations (where the connection $\Gamma^{c}_{ab}$ is 
independent of the metric at the level of the action) or purely affine
theories \cite{BanadosFerreira2010}, provided a metric can be defined. 
\item {\it The field equations are at most 2$^{\rm nd}$ order.} This does restrict the class of
acceptable theories (for example $F(R)$ theories have generally higher derivatives) but it
has become clear that it is these higher order terms that lead (again) to instabilities in
the generation of large scale structure.
\item {\it The field equations are gauge-form invariant.}  Gauge form-invariance is the linearized version of the full
diffeomorphism invariance of any gravitational theory with a manifold structure. It is the unbroken symmetry of the field equations under
gauge transformations. After a gauge transformation, the field equations retain their exact form : they are {\emph form-invariant}
(see ~\cite{Straumann2004,Giulini2006,WestmanSonego2007a,WestmanSonego2007b} for further discussion).
\end{enumerate}
It is of course possible to relax some of these conditions and we will discuss how in the
conclusions.

Armed with these conditions we can construct $U$ in Fourier space solely 
out of $\Phi$ and ${\dot \Phi}$ such that:
$U_{\Delta}= k^2 A\Phi$, $U_{\theta}= k B\Phi$, $U_{P}= k^2 C_1\Phi + kC_2{\dot \Phi}$ and
$U_{\Sigma}=D_1\Phi+D_2{\dot \Phi}/k$. All operators above  (e.g. $A$ ) are dimensionless.
Terms such as $\ddot{\Phi}$, $\dot{\Psi}$ and $\ddot{\Psi}$ (and higher derivatives) are forbidden while terms proportional to $\Psi$ are allowed but their
coefficients vanish as a result of the Bianchi identities. This last fact is a consequence of the consistency conditions outlined above
and the reader is referred to  \cite{Skordis2008b} for a more thorough treatment.

We can rearrange (\ref{NP1}) in the form of (\ref{eq_eff_Poisson}) to immeadiately read off
\begin{eqnarray}
\mu^2=1+\frac{A+3 \adotoa_k B}{2} \nonumber.
\end{eqnarray}
where we let $\adotoa_k \equiv \adotoa/k$.

Furthermore we have that the shear field equation becomes
\begin{eqnarray}
\Phi-\Psi&=&\zeta \Phi+  \frac{g}{k} {\dot \Phi} \nonumber
\end{eqnarray}
where we have let $\zeta = D_1$ and $g = D_2$.

We see that  the gravitational slip $\Phi - \Psi$ will generally depend on both $\Phi$ \emph{and its time derivative},  $\dot{\Phi}$,
and not solely on $\Phi$.
Defining $\Phi = \gamma_{PPF} \Psi $ as has been in proposed in \cite{DanielEtAl2009,DanielEtAl2010},  we find that 
\begin{eqnarray}
\gamma_{PPF}[\Phi] = \frac{1}{1 - \zeta - \frac{g}{k}  \frac{d\ln \Phi}{d\tau} } \approx 1 + \zeta +  \frac{g}{k}  \frac{d\ln \Phi}{d\tau}
\end{eqnarray} 
Thus unless $g=0$, $\gamma_{PPF}$ is an explicit functional of $\Phi$, introducing interesting enviromental dependance on
the matter distribution.  All parameterizations of the slip used so far, for which $\Phi - \Psi \propto \Phi$, have ignored this
possibility  which suggests that they were over simplistic (although see \cite{HuSawicki2007a}).

As shown in \cite{Skordis2008b}, as a result of the Bianchi identities we have $\nabla_a U^a_{\phantom{a}b} = 0 $,
 leading to a series of restrictions on the coefficients:
\begin{eqnarray}
 A &=& \adotoa \frac{2 \adotoa_k \dot{g} +  2 k (2\adotoa_k^2 + \frac{1}{3})g - 2 \adotoa \zeta}{ \dot{\adotoa} - \adotoa^2 - \frac{k^2}{3}  }
  \nonumber 
\\
B &=& - \frac{k}{3\adotoa} A  - \frac{2}{3} g
  \nonumber 
\\
  C_1  &=& \frac{3}{k} ( \dot{B} + 2 \adotoa  B  ) + 2 \zeta = -A - \frac{1}{\adotoa} [\dot{A} + k B ]
\nonumber 
\\
C_2&=&-\frac{k}{\adotoa} A \label{consistency}
\end{eqnarray}
This means a  consistent modification to the Einstein equations
is uniquely determined by two arbitrary free functions, $\zeta(\tau,k)$ and $g(\tau,k)$.

Finally we can combine the expression for $A$ and $B$ appropriately to find  that $g$ 
has a simple interpretation as a perturbation of the effective gravitational constant :
\begin{eqnarray}
\mu^2 = 1 - \frac{\adotoa}{k} g
\end{eqnarray}
Hence the consistency conditions lead to an important relationship between a generalized form
of the gravitational slip. In particular, {\it if we consider time variations of the Newtonian potential, it
is inconsistent to consider a restricted parametrization of the gravitational slip of the form $\Phi-\Psi=\zeta\Phi$ on all scales.}~\footnote{This
may be acceptable  on small scales, however,  where the additional contributions to $\Phi-\Psi$ may be recasted in terms of $\Phi$, thus, creating
an effective $\Phi - \Psi = \zeta_{eff} \Phi $}.

\subsection{Parameterizing $\zeta$ and $g$}
\label{zeta_g_param}

We are not assuming any particular underlying theory of modified gravity and hence do not
have a specific model for $\zeta$ and $g$. Our interest is in theories that may mimic the behaviour
of dark energy so we expect deviations from Einstein gravity to emerge as $\Omega_E$ begins
to diverge from $0$. A simple assumption is to assume that the gravitational slip is analytic at $\Omega_E=0$
and Taylor expand it:
\begin{eqnarray}
\zeta=\zeta_1\Omega_E+\zeta_2\Omega_E^2+{\cal O}(\Omega_E^3) \label{smallp1}
\end{eqnarray}
We can do the same for $g$ so that 
\begin{eqnarray}
g=g_1\Omega_E+g_2\Omega_E^2+{\cal O}(\Omega_E^3) \label{smallp2}
\end{eqnarray}
In Section \ref{small} we find how the growth depends on such a parametrization and, in particular,
determine analytic expressions for $\gamma$.

This way of parameterizing $\zeta$ and $g$ has three major advantages : 
\begin{itemize}
\item It is in the spirit of the  Parametrized Post-Newtonian (PPN) formalism where 
the PPN parameters are isolated from the potentials which are dependent on the density profiles
and thus the solutions; the role of the "potential" in this case is taken by $\Omega_E(\tau)$ which depends on the background cosmology. 
\item Expanding in powers of $\Omega_E$ isolates the 
background effects of the dark energy from the genuine effects of the perturbations. 
In particular, the dark energy relative density $\Omega_{0E}$ or the dark matter relative density $\Omega_{0m}$ would have no effect on
the parameters $\zeta_i$ and $g_i$.
\item This expansion makes mathematical sense for any analytic function,
as the function $\Omega_E(\tau)$ is always bounded to be $0\le \Omega_E \le 1$, i.e. it is a naturaly small parameter.
\end{itemize}

Note that we have dropped any $k$ dependence from this parametrization. There are two ways that $k$-dependence can enter,
either relative to a fixed scale $k_0$ (which may be part of some theory of gravity) or relative to the Hubble scale $\adotoa$.
 If we wish to see how our results are affected by a scale dependence relative to the temporal changes introduced by the FRW  background
 we can extend the parametrization to
\begin{eqnarray}
\zeta=\zeta^{(0)}+\zeta^{(1)}{\cal H}_k \nonumber \\
g=g^{(0)}+g^{(1)}{\cal H}_k \label{largescale}
 \end{eqnarray}
where, as above, we have
\begin{eqnarray}
\zeta^{(0)}&=&\zeta_{01}\Omega_E+{\cal O}(\Omega_E^2) \nonumber \\
\zeta^{(1)}&=&\zeta_{11}\Omega_E+{\cal O}(\Omega_E^2) \label{zeta_adotoa_exp}
\end{eqnarray}
and likewise with $g^{(0)}$ and $g^{(1)}$.

It turns out that, if we attempt to, on one hand generalize our parametrization of $\zeta$ and $g$, but, 
on the other hand pin it down so as to be consistent with PPN method 
used on much smaller scales, we need to change our previous approach. 
It is entirely possible that there are other scales in the system. Furthermore, as we show in
in Appendix \ref{app_PPN}, to leading order we may have $g\propto {\cal H}_k^m$ where 
$m$ is negative. These can complicate the simple model we considered above.
 For example, consider the function $f =e^{\ell \adotoa}$ where $\ell$ is a fixed scale.
How does one expand this on small scales? One would be tempted to write $f = e^{\ell k \adotoa_k} \approx 1 + \ell k \adotoa_k = 1 + \ell \adotoa $ but this
clearly makes no sense. The scale $k$ was artificially introduced and leads to erroneous conclusions.

 In general we have a function $\zeta(\tau,k)$. Since $\zeta$ is dimensionless while $\tau$ and $k$ are not, the functional dependence on $\tau$ and $k$ must
come in dimensionless combinations. It is convinient to exchange $\tau$ with either $\tau(\adotoa)$ or with $\tau(\Omega_E)$. Thus the most general 
function $\zeta$ will have the form $\zeta = \zeta(\Omega_E, \adotoa_k, \ell\adotoa, k/k_0)$ for constants $\ell$ and $k_0$, and there may be additional
dimensionless parameters entering. We can thus isolate the leading-order dependence of $\zeta$ on $\adotoa_k$ and write
\begin{equation}
\zeta = \zeta_L(\Omega_E,k) \adotoa_k^n
\end{equation}
for a constant $n$.
We expand $g$ in a similar way as
\begin{equation}
g = g_L(\Omega_E,k) \adotoa_k^m
\end{equation}
for a constant $m$.
 In Appendix \ref{app_PPN} we show that a consistent PPN limit fixes $n=0$ and $m=-1$.

Thus we arrive at our general expansion of $\zeta$ and $g$ in the small scale limit, which is consistent with PPN:
\begin{eqnarray}
\zeta &=& \zeta^{(0)} \nonumber
\\
g &=& g^{(-1)} \frac{1}{\adotoa_k} \nonumber
\\
\zeta^{(0)} &=& \zeta_1\Omega_E+\zeta_2\Omega_E^2+\zeta_3\Omega_E^3 \label{zeta_param}
\\
g^{(-1)} &=& g_1\Omega_E+g_2\Omega_E^2+g_3\Omega_E^3\label{g_param}
\end{eqnarray}
The parameters $\zeta_i$ and $g_i$ may in principle be $k$-dependent, e.g. $\zeta_1 \propto (k/k_0)^N$ for a fixed scale $k_0$ and power index $N$. We shall not
investigate this further in this work but we stress it as a possibility and note that our results would include these cases.
In Section \ref{General} we conclude by presenting and analysing the resulting growth rate due to such a parametrization.

\section{The growth rate on small scales for a simplified model of $\zeta$ and $g$}
\label{small}
The definition of $\gamma$ originally arose when characterizing the evolution of
small scale density perturbations. We expect it to be particularly useful when characterizing
the growth of structure on small scales (by which we mean roughly between 1 and 200$h^{-1}$ Mpc) as would be probed by galaxy redshift surveys (through
redshift measurements of the power spectrum, for example, or redshift space distortions \cite{Kaiser1987,GuzzoEtAl2008,Simpson10}) and weak lensing surveys.

 In this section we focus on the behaviour of this system in the limit in which $\adotoa_k  \equiv \adotoa/k \ll 1$, i.e. on
scales deep inside the horizon. We can then assume that $\Delta_{DE}\simeq\theta_{DE}\simeq0$. This is true for $c_s^2 \sim O(1)$ or larger.
Since in this paper we are concerned with modifications of gravity rather than the speed of sound we will
leave the full treatment of small $c_s^2$ for a future investigation.  We shall, however, show the effect of small $c_s^2$ numerically futher below.

In what follows we will present a modified evolution equation, find analytic expressions for
the growth factor and compare to numerical results for the full system.

\subsection{Evolution of density perturbations}
 Combining the fluid equations (\ref{fluid}) in one 2$^{\rm nd}$ order equation, we find that $\Delta_M$ obeys 
\begin{eqnarray}
{\ddot \Delta}_M+ \adotoa \; {\cal U} \, {\dot \Delta}_M-\frac{3}{2} \Omega_M \adotoa^2  \;  {\cal V}{\Delta}_M&=&0 \label{largek}
\end{eqnarray}
with the damping coefficient modified by
\begin{eqnarray}
{\cal U}[\zeta,g] &=& 1 +  \frac{3\Omega_M \adotoa_k }{2(1 - \adotoa_k g ) }\bigg\{     g  
\nonumber 
\\
&& \ \ \ \ + \frac{3\adotoa_k \left[ 1-\zeta  -  g  B/2 \right]}{1 - \adotoa_k g +9\adotoa^2_k\Omega_M/2}\bigg\}
\end{eqnarray}
and the response term modified by.
\begin{eqnarray}
{\cal V}[\zeta,g] &=&  \frac{1 -  \zeta -gB/2}{(1- \adotoa_k g )(1 + 9\adotoa_k^2\Omega_M/2 - \adotoa_k g ) }
\end{eqnarray}
Specifying $\zeta$ and $g$ completely fixes $\cal{ U}$ and $\cal{ V}$.

If we further take the provisional small scale limit $\adotoa_k \ll 1$ (i.e. without assuming anything about $\zeta$ and $g$) we find that
\begin{eqnarray}
{\cal U}[\zeta,g] &=& 1 +  \frac{3\Omega_M \adotoa_k }{2(1 - \adotoa_k g ) }\bigg\{     g  
 + \frac{3\adotoa_k \left[ 1-\zeta  -  g  B/2 \right]}{1  - \adotoa_k g}\bigg\} \nonumber \\
 \label{smallU}
\end{eqnarray}
and
\begin{eqnarray}
{\cal V}[\zeta,g] &=&  \frac{1 -  \zeta -gB/2}{(1-\adotoa_k g)^2} \label{smallV}
\end{eqnarray}
The full  small scale limit, including $\zeta$ and $g$ is presented in appendix \ref{app_PPN}.

\subsection{Analytic expressions for the growing mode}
\label{analytic_simple}
We expect modifications of gravity to kick in when the expansion rate starts to deviate from
matter domination. In this section we will work with the parametrization of $\zeta$ and $g$
proposed in equations (\ref{smallp1}) and (\ref{smallp2}). We can immediately see from equations (\ref{smallU}) and (\ref{smallV}) that
the effects from $g$ will only come in at order ${\cal H}_k$.
We shall also restrict ourselves to constant $w$ and leave varying $w$ for section \ref{General}. In this section we shall present the derivation
and result to first order in $\Omega_E$ and then present the result to second order in 
$\Omega_E$. 

The starting point is 
\begin{eqnarray}
\ddot{\Delta}_M +   \adotoa \dot{\Delta}_M
 -\frac{3 \adotoa^2  \Omega_M }{2} (1 - \zeta_1\Omega_E) \Delta_M = 0 \nonumber
\end{eqnarray}
Changing variables to $\ln a$ and defining $\Delta_M = a Y$, we can rewrite this equation as
\begin{eqnarray}
 Y'' 
+ \frac{5- 3w  \Omega_E  }{2}  Y'
+ \frac{3}{2} \Omega_E [1 - w
 +   \zeta_1 \Omega_M  
] Y = 0 \nonumber
\end{eqnarray}
where we have used the Raychauduri equation  (\ref{Ray}) and where we set $()' = \frac{d}{d\ln a}$.

Changing variables to $\Omega_E$ and using the 0$^{\rm th}$ order fluid
conservation equation, rewritten as $\Omega_E' =  -3w\Omega_M  \Omega_E$
we find
\begin{eqnarray}
3 w^2\Omega_M^2 \Omega_E^2\frac{d^2Y}{d\Omega_E^2}  
&+& \frac{w}{2}   \Omega_M \Omega_E  \left[ 
 3w( 2  -3\Omega_E)  
- 5 
 \right] \frac{dY}{d\Omega_E} \nonumber \\
&+& \frac{1}{2}  [ \Omega_E (1 - w)
 +   \zeta_1 \Omega_M  \Omega_E
] Y = 0 \label{hyper}
\end{eqnarray}

Equation (\ref{hyper}) has a three regular singular points (at $\Omega_E = 0$  and $\Omega_E = 1$) and can therefore
be transformed into the  hypergeometric equation.  We wish to find its behaviour around $\Omega_E=0$ and
do so by expanding $Y$,  $Y = 1 + Y_1 \Omega_E$ to find (to lowest order in $\Omega_E$)
\begin{eqnarray}
 Y_1 =  \frac{  1 - w +     \zeta_{1}  }{w(5-6w)} \nonumber
\end{eqnarray}
and hence
\begin{eqnarray}
\Delta_M = a \left[ 1 +   \frac{  1 - w +     \zeta_{1}  }{w(5-6w)}  \Omega_E\right] \nonumber
\end{eqnarray}
Therefore to ${\cal O}(\Omega_E)$ 
\begin{eqnarray}
\ln \Delta_M =  \ln a  +   \frac{1- w  +  \zeta_{1}  }{w(5-6w)}  \Omega_E \nonumber 
\end{eqnarray}
which we can use to find the logarithmic derivative of the growth factor $f\equiv d\ln\Delta_M/d\ln a$:
\begin{eqnarray}
f =  1 - \frac{3(1 - w + \zeta_{1}) }{5-6w} \Omega_E \nonumber
\end{eqnarray}
As stated above, we are parametrizing the growth factor using $f=\Omega_M^\gamma$
and
so we have that
\begin{eqnarray}
\gamma =  \gamma_0  =    \frac{3(1 - w + \zeta_{1}) }{5-6w}  \label{0gamma}
\end{eqnarray}
where the subsciript $0$ is due to the fact that this is the lowest order approximation.

We can easily see that, for $w=-1$ and $\zeta_1=0$ we retrieve $\gamma=6/11\simeq 0.54545...$,
the approximation first proposed in \cite{Peebles1980} and subsequently rederived and
advocated in \cite{WangSteinhardt1998} and \cite{Linder2005a}. If we assume a more general (but constant)
equation of state, we improve on the approximation advocated in \cite{Linder2005a}. 

It is possible to further improve the approximation by going to next order in $\Omega_E$. For this we need $Y$ and $\zeta$ to ${\cal O}(\Omega_E^2)$,
i.e. $Y = 1 + Y_1 \Omega_E + Y_2 \Omega_E^2$ and  $\zeta = \zeta_1 \Omega_E + \zeta_2 \Omega_E^2$.
. We now have that
\begin{eqnarray}
\gamma=\gamma_0+\gamma_1\Omega_E
\end{eqnarray}
where $\gamma_0$ is as derived above and given in (\ref{0gamma}) and
\begin{eqnarray}
\gamma_1 &=& \frac{3w}{2}  \left[  - 3Y_1   -(2 - 3w) Y_1^2 + 4 Y_2 \right] \label{highernohk}
\end{eqnarray}
where we have defined
\begin{eqnarray}
Y_2 &=& \frac{ ( 1 - w  )  ( 15 w^2   - 4 w -  1 ) + \zeta_{1}  ( 9 w^2   + 2 w -  2 )} 
{2w^2(12w-5)(5-6w)}
\nonumber \\ & & \ \ \ \ - \frac{\zeta_{1}^2 + w (5-6w) \zeta_{2} }{2w^2(12w-5)(5-6w)}  
\end{eqnarray}
In the case of $w=-1$ (i.e. the cosmological constant) and $\zeta=0$ we find that expression (\ref{highernohk}) reduces to
\begin{eqnarray}
\gamma_1=\frac{15}{2057} \nonumber
\end{eqnarray}

As stated above assuming that $g$ can be parametrized in the same way as $\zeta$, independently of $\adotoa_k$, we find that it does not affect the growth of structure on small scales.
The situation is of course different if we consider expanding $1-\mu^2=g\adotoa_k$ in powers
of $\Omega_E$ with coefficients which are ${\cal H}_k$ independent.  

\subsection{Comparison with numerical results}
\label{first_numerics}
\begin{figure}[htbp]
\begin{flushleft}
\vspace{-15pt}
\epsfig{figure=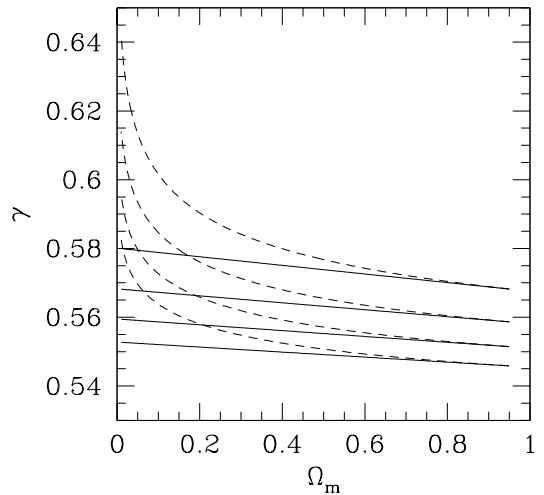,width=9cm}
\end{flushleft}
\vspace{-60pt}
\caption{The growth parameter, $\gamma$, for a selection of dark energy models, as
a function of $\Omega_M$. The dashed curves are the numerical results for
 $w=-1$, $-0.8$, $-0.6$ and $-0.4$ in ascending order and the corresponding analytic approximations are plotted in solid line.}
\label{justDE}
\vspace{-1pt} 
\end{figure}

We can solve (\ref{field}), (\ref{fluid}) and (\ref{consistency}) to assess the quality of
this analytic approximation. We first restrict ourselves to a dark energy like fluid with
a large sound speed, $c_s^2 \sim O(1)$, (such as a quintessence model or most other field like models)
and assume no modifications to gravity. We see  a number of features in Figure \ref{justDE}. First of
all, $\gamma$ is very clearly not independent of $\Omega_M$ as has been generally
assumed. In fact as $\Omega_M\rightarrow 0$, $\gamma$ deviates substantially from
its asymptotic value at $\Omega_M=1$. Nevertheless, we find that (\ref{highernohk}) is a good aproximation to the true behaviour. In Figure
\ref{justDE} we plot the true and approximate behaviours  of $\gamma$ for
$w=-1$, $-0.8$, $-0.6$ and $-0.4$: we find deviations
of at most a ten percent for $w=-0.4$ at  $\Omega_M=0.1$. 

\begin{figure}[htbp]
\begin{flushleft}
\vspace{-15pt}
\epsfig{figure=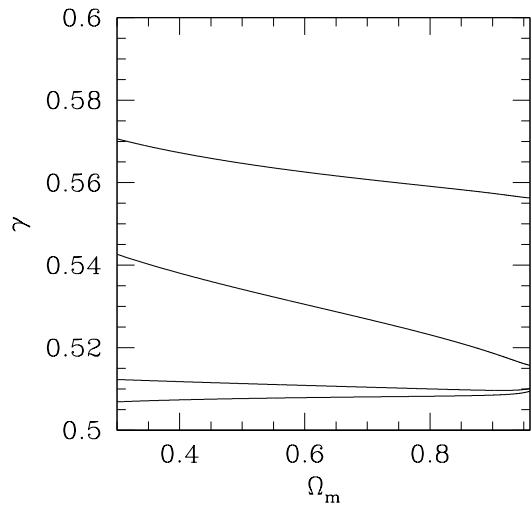,width=9cm}
\end{flushleft}
\vspace{-60pt}
\caption{The growth parameter, $\gamma$, for a selection of dark energy models where $w=-0.6$ and the sound speed is chosen to be $c_s^2=5\times 10^{-4}$,
 $c_s^2=10^{-5}$, $c_s^2=10^{-6}$ and $c_s^2=0$ in descending order from the top of the figure, as
a function of $\Omega_M$.}
\label{justDECS}
\vspace{-15pt} 
\end{figure}

We may relax the condition $c_s^2=1$ substantially before the dark energy perturbations
affect the growing mode in the density field. This is clearly illustrated in Figure \ref{justDECS}
where, for $w=-0.6$, three different values of the sound speed are chosen: $c_s^2=5\times 10^{-4}$, $c_s^2=10^{-5}$
 $c_s^2=10^{-6}$ and  $c_s^2=0$. For $c_s^2=5\times10^{-4}$, the growth is still indistinguishable from
$c_s^2=1$ and only once the jean scale for the dark energy falls susbtantially below
the cosmological horizon is the effect noticeable.  In particular our results hold for values $c_s$ such that $c_s > 10 \sim \adotoa_k$,
which approximately translates to $c_s^2 > \sim 3 \times 10^{-4}$.

We note in passing that the effects of the speed of sound without modifications of gravity, have been studied 
in \cite{WellerLewis2003,BeanDore2004,BallesterosRiotto2008}. 
In particular \cite{BallesterosRiotto2008} 
have found fitting formulas for $\gamma$ which interpolate between $c_s=0$ and $c_s \sim O(1)$. However, those fitting formulas do not account for
modifications of gravity and are in fact quite model dependent (they depend on the background cosmology).
The current constraints on the speed of sound do not rule out
small values~\cite{dePutterHutererLinder2010} and in fact they are consistent with $c_s=0$. 
Since this work  concerns the effects of modifications of gravity rather than effects coming from the speed of sound, however,
we leave the case for small $c_s$ for a future investigation.

\begin{figure}[htbp]
\begin{flushleft}
\vspace{-15pt}
\epsfig{figure=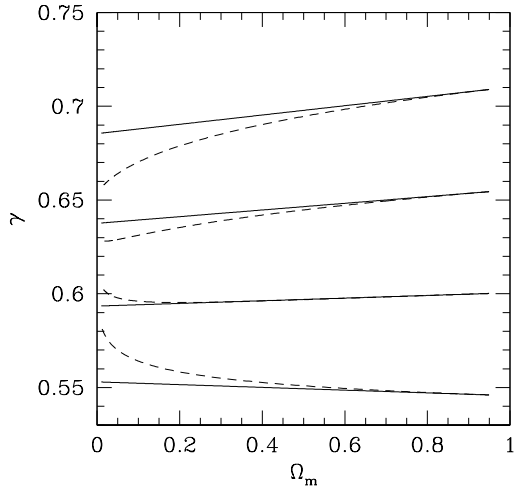,width=9cm}
\end{flushleft}
\vspace{-60pt}
\caption{The growth parameter, $\gamma$, for a selection of gravitational
slip parameters of the form $\zeta=\zeta_1\Omega_E$, as
a function of $\Omega_M$. The dashed curves are the numerical results for
 $\zeta_1=0$, $0.2$, $0.4$ and $0.6$ in ascending order and the corresponding analytic approximations are plotted in solid line.}
\label{slipno2}
\vspace{-1pt} 
\end{figure}

\begin{figure}[htbp]
\begin{flushleft}
\vspace{-15pt}
\epsfig{figure=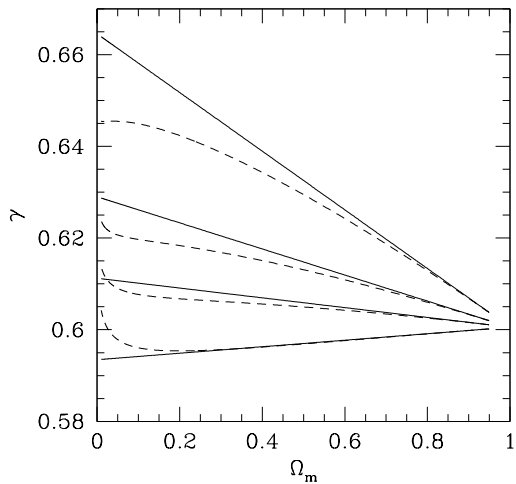,width=9cm}
\end{flushleft}
\vspace{-60pt}
\caption{The growth parameter, $\gamma$, for a selection of gravitational
slip parameters of the form $\zeta=\zeta_1\Omega_E+\zeta_2\Omega_E^2$, as
a function of $\Omega_M$. The dashed curves are the numerical results for $\zeta_1=0.2$ and 
 $\zeta_2=0$, $0.125$, $0.2$ and $0.4$ and the corresponding analytic approximations are plotted in solid line.}
\label{slip2}
\vspace{-1pt} 
\end{figure}

Let us now introduce modifications to gravity and assume a non-negligible gravitational slip.
In Figures \ref{slipno2} and \ref{slip2} we show how well the analytic approximation fares in
comparison to different values of $\zeta_1$ and $\zeta_2$ ( in these figures we restrict ourselves
to $w=-1$ but the agreement between the numerical and approximate estimates of $\gamma$ is
generic). In Figure \ref{slipno2} we restrict ourselves to a gravitational
slip which is linear in $\Omega_E$ and identify the two main effects. First of all, the quicker
the onset of slip, the more effective the supression of growth due to the onset of dark energy-
$\gamma_0$ increases with $\zeta_1$. Furthermore, the dependence of $\gamma$ on $\Omega_E$, through 
$\gamma_1$, changes sign so that for larger $\zeta_1$, $\gamma$ becames smaller as $\Omega_M$ decreases.

This last effect is further affected by $\zeta_2$. Indeed we find that the slope of $\gamma$ as
function of $\Omega_M$ can greatly be affected by higher order terms in $\zeta$.

\subsection{Intermediate and large scales}
\label{large}
While the $\gamma$ parametrization is particularly useful on small scales where terms
dependent on ${\cal H}_k$ can be discarded, this isn't true once we look at horizon crossing,
i.e. ${\cal H}_k\simeq 1$. This regime will be of particular importance for measurements of
the cosmic microwave background and using the Integrated Sachs-Wolfe effect to look
for the presence of dark energy or modified gravity \cite{CrittendenTurok1996,Hu2008,DanielEtAl2009,DanielEtAl2010}. 
Let us now consider the parametrization presented in equations (\ref{largescale})

We expand eq. (\ref{largek}) to include the first order term in ${\cal H}_k$ which gives
\begin{eqnarray}
&&\ddot{\Delta}_M + \adotoa \left[ 1 +    \frac{  3  \adotoa_k\Omega_M }{2}  g^{(0)} \right] \dot{\Delta}_M 
- \frac{3}{2}\Omega_M \adotoa^2 \bigg\{ 1- \zeta^{(0)}  
\nonumber \\
&&
 \ \ \ \ 
+ \adotoa_k \left[ g^{(0)} \left(2   -  
\zeta^{(0)}\right) - \zeta^{(1)}   \right]
 \bigg\} \Delta_M = 0 
\end{eqnarray}
where the functions $\zeta^{(0)}$,  $\zeta^{(1)}$and   $g^{(0)}$ are further expanded in powers of $\Omega_E$ using equation (\ref{zeta_adotoa_exp}).
We apply the same techniques as in section \ref{analytic_simple} above, and expand $\gamma$ as
\begin{eqnarray}
\gamma=\gamma^{(0)}+\gamma^{(1)}{\cal H}_k \nonumber 
 \end{eqnarray}
where the coefficients $\gamma^{(0)}$ and $\gamma^{(1)}$ are further expanded in powers of $\Omega_E$, i.e. $\gamma^{(0)} = \gamma_{00} + O(\Omega_E)$.
Carrying through the expansion we find that
\begin{eqnarray}
\gamma^{(0)} &=&   3  \frac{  1 - w +     \zeta_{01}  }{5-6w} 
 + O(\Omega_E)
\\
\gamma^{(1)} &=&  \frac{27}{4} \; \frac{3 \zeta_{01}^2 + 2\zeta_{11}   -2 g_{01}}{3w-1} 
 + O(\Omega_E)
\end{eqnarray}
There are now $3$ free modified gravity constants: $\zeta_{01}$, $\zeta_{11}$ and $g_{01}$.
Notice how the first order correction in $\adotoa_k$ to $\gamma$  only depends on the modified gravity parameters and is zero
 for $wCDM$, i.e. corrections in  $\adotoa_k$ to $\gamma$  for $wCDM$ come to second order.

We find that scale dependent corrections in powers of $\adotoa_k$ are always subdominant compared with corrections in powers of $\Omega_E$,
 or corrections with respect to a fixed scale, e.g. $\sim (\ell k)^N$. They are thus effectively negligible. 
We find that at $k\sim 0.04  h Mpc^{-1}$ corrections  in powers of $\adotoa_k$  are around $1\%$ at redshift $z=1$ for $\zeta_{11} = -0.6$
and become smaller at larger $k$ (smaller scales), or lower redshift ($z\sim 0$).
Hence, it perfectly reasonable to discard scale dependent corrections which come in powers of $\adotoa_k$.

\section{General evolution of $\zeta$ and $g$}
\label{General}
We now wish to address a more realistic expansion of the $\zeta$ and $g$ proposed in the Section \ref{formalism} which is consistent 
with the PPN approximation on small scales, namely (\ref{zeta_param}) and (\ref{g_param}). 
Having in mind our findings in section \ref{large} 
on intermediate scales, we disregard any dependence of $\Delta_M$ (and hence of $\gamma$) on $\adotoa_k$, and therefore
to this order we can set $\Delta_M = \delta_M$.

 We perform the calculation in steps.
First, we solve the 2nd order differential equation obeyed by $\delta$ by applying a Taylor series expansion
in $\Omega_E$, resulting in a set of coefficients $\{Y_1,Y_2,Y_3\}$ which are functions of
the expansion coefficients of $w$, $\Uc$ and $\Vc$.
  Then we relate the a perturbative expansion coefficients  for $\delta$, namely  $\{Y_1,Y_2,Y_3\}$, to the $\gamma$ parameter.
Finally we relate  $\{Y_1,Y_2,Y_3\}$ for general $\Uc$ and $\Vc$ to the specific case of our $g\zeta$CDM model.

\subsection{Perturbative solution of the $\delta$-equation}
\label{general_delta}

We start with the equation for the matter density contrast in the absence of DE perturbations, namely
\begin{equation}
\ddot{\delta}_M +   \adotoa \Uc \dot{\delta}_M -\frac{3 \adotoa^2  \Omega_M }{2} \Vc \delta_M = 0 
\label{eq_delta_M_app}
\end{equation}
Changing the independent variable from $\tau$ to $\ln a$ and the dependent variable from $\delta_M$ to $Y$ defined by $Y \equiv\delta_M/a$, we get
\begin{eqnarray}
&&
 Y''
+ \left[ \Uc + \frac{3}{2}(1-w\Omega_E) \right]Y'
\nonumber 
\\
&&
\ \ \ \
+ \left[ \Uc + \frac{1}{2}(1-3w\Omega_E) -\frac{3   \Omega_M }{2} \Vc \right]Y
= 0 
\label{eq_Y_app}
\end{eqnarray}
On small scales we may expand $Y(k,\tau) = Y^{(0)}(k,\tau) +  Y^{(1)}(k,\tau) \adotoa_k + O(\adotoa_k^2)$ (see appendix \ref{app_PPN}),
where the functional coefficients $ Y^{(i)}(k,\tau)$  have no dependence on $\adotoa_k$ but may still be $k$ or $\tau$ dependent through
combinations of the form $\ell k$ or $\tau/\ell$ (where $\ell$ is some scale, not necessarily the same scale for all such combinations). We are interested
in the small-scale limit $\adotoa_k\rightarrow 0$, and we shall work with  $Y(k,\tau) = Y^{(0)}(\ell k,\Omega_E)$ only. As discussed in section \ref{large} above,
at the scale of validity of the $\gamma$ parameterization $\adotoa_k$-corrections are always small and irrelevant.
Since time dependence only comes through  $\tau/\ell$  for some scale $\ell$ we may further exchange $\tau/\ell$ with a function of $\Omega_E$, thus we let
 $Y(k,\tau) = Y(k,\Omega_E)$.

We now Taylor expand $ Y(k,\Omega_E)$ in powers of $\Omega_E$.
 To get $\gamma$ to $O(\Omega_E^2)$  we need to expand $Y$ to $O(\Omega_E^3)$ as
\begin{equation}
Y = 1 + Y_1 \Omega_E + Y_2 \Omega_E^2 + Y_3 \Omega_E^3
\label{eq_Y_taylor}
\end{equation}
We then use the above expansion (\ref{eq_Y_taylor}) into (\ref{eq_Y_app}) and match orders~\footnote{We could have used a more general form $Y(0) = Y_0$, however,
$Y_0$ can always be absorved into an overall normalization of $\delta_M$. }. To be able to do that we need to expand the functions $w$, $\Uc$ and $\Vc$. 
Since $w$ always appears in the combination $w \Omega_E$ we only need it to $O(\Omega_E^2)$. The functions $\Uc$ and $\Vc$, however, are needed to $O(\Omega_E^3)$.
Thus  we expand
\begin{eqnarray}
w &=& w_0 + w_1 \Omega_E + w_2 \Omega_E^2 \label {w_param}
\label{app_w_exp}
\\
\Uc &=& 1 + \Uc_1 \Omega_E +  \Uc_2 \Omega_E^2 +  \Uc_3 \Omega_E^3  
\label{app_U_exp}
\\
\Vc &=& 1 + \Vc_1 \Omega_E +  \Vc_2 \Omega_E^2 +  \Vc_3 \Omega_E^3  
\label{app_V_exp}
\end{eqnarray}
While $w_i$ are constants, the $\Uc_i$ and $\Vc_i$ coefficients may be $k$-dependent, for example $\Uc_1 = \Uc_{01} (\frac{k}{k_0})^N$ for some index $N$ and scale $k_0$.
Using the expansions (\ref{eq_Y_taylor}), (\ref{app_w_exp}),  (\ref{app_U_exp}) and (\ref{app_V_exp}) into (\ref{eq_Y_app}) and equating orders in $\Omega_E$
we find 
\begin{eqnarray}
  Y_1 &=& \frac{ -1 + w_0 + \Vc_1 - \frac{2}{3}\Uc_1  }{ w_0(6w_0 - 5) }
\nonumber 
\\
Y_2 
&=&
 \frac{1}{2w_0(12 w_0 - 5)}\bigg\{
 w_1 +  \Vc_2 -  \Vc_1 -\frac{2}{3} \Uc_2 
\nonumber 
\\
&&
+\bigg[-1 -\frac{2}{3} \Uc_1 +  \Vc_1 + 5 w_1 
\nonumber 
\\
&&
 +w_0 (- 4  + 2 \Uc_1 + 15 w_0 -18 w_1 ) \bigg] Y_1
\bigg\}
\nonumber 
\\
Y_3 &=&
\frac{1}{3w_0(18 w_0 - 5)}\bigg\{ \Vc_3 -\frac{2}{3} \Uc_3 - \Vc_2 + w_2
\nonumber
\\
&&
+\bigg[ \Vc_2 -\frac{2}{3}\Uc_2 -\Vc_1 -4w_1 +5w_2 +2w_1 \left( \Uc_1- 6 w_1\right)
\nonumber
\\
&&
+ w_0 \left( -2 \Uc_1 + 2 \Uc_2 -9 w_0 + 42 w_1 -24  w_2 \right) \bigg] Y_1
\nonumber
\\
&&
+ \bigg[ -1 -\frac{2}{3} \Uc_1 +  \Vc_1 -9 w_0 + 10 w_1
\nonumber
\\
&&
  + 2w_0\left( 2\Uc_1  + 27w_0  -30  w_1\right)\bigg]Y_2 \bigg\}
\label{Ytot_param}
\end{eqnarray}
We notice that in general there are  $9$ initial coefficients appearing in (\ref{eq_delta_M_app}) 
that determine only $3$ final coefficients  $Y_i$ for the solution to (\ref{eq_delta_M_app}).

Having found the coefficients $Y_i$ we proceed to relate them to $\gamma$.

\subsection{From $\delta$ to $\gamma$}

We can now  use the definition of the logarithmic change of the growth-rate $f = \frac{d\ln \delta}{d\ln a}$ and then
  get $\gamma$ from $\gamma = \ln f/\ln \Omega_M$.  We find that $\gamma$ is expanded as
\begin{equation}
\gamma = \gamma_0 + \gamma_1 \Omega_E + \gamma_2 \Omega_E^2  \label{gamma_param}
\end{equation}
where
\begin{eqnarray}
\gamma_0 &=& 3 w_0 Y_1
\nonumber 
\\
\gamma_1 &=&
- \gamma_0 \left(\frac{3}{2} +   Y_1  - \frac{1}{2}  \gamma_0     \right)
+ 3 w_1  Y_1
+ 6 w_0 Y_2
\nonumber 
\\
\gamma_2
&=&
 \gamma_0 \left( - \frac{11}{6}   -\frac{1}{2} \gamma_0 +\frac{1}{3} \gamma_0^2 \right) - \frac{3}{2}  \gamma_1
\nonumber 
\\
&&
+ \frac{1}{2} \left[ 6 w_1 \gamma_0 +6 w_2 - \gamma_0^2 -3\gamma_0  -2\gamma_1 \right] Y_1
\nonumber 
\\
&&
+  \left[ -(1 - 6w_0)\gamma_0 +6 w_1  \right] Y_2 + 9 w_0 Y_3
\label{gammatot_param}
\end{eqnarray}

Given a set of coefficients $\{Y_1,Y_2,Y_3,w_0,w_1,w_2\}$ we can get $\gamma(\Omega_E)$.
Note that  $\{Y_1,Y_2,Y_3\}$ may be $k$-dependent, for example $Y_1 = Y_{01} (\frac{k}{k_0})^N$ for some index $N$ and scale $k_0$.

One important point is in order. What we have done so far is more general than the approach we discussed in the main part of the article.
In particular the derivation of the $\gamma$ coefficients in this appendix would hold for any theory for which the density contrast obeys
(\ref{eq_delta_M_app}). One such theory is DGP, even though strictly speaking DGP does not fit within our framework of the main part of the article.

To connect the $\gamma$ coefficients above with our framework we must perform a third step : relate the $\Uc_i$ and $\Vc_i$ coefficients with expansions of 
$g$ and $\zeta$.

\subsection{Relating to the  $g\zeta$CDM model}
As shown in the appendix \ref{app_PPN}, to be consistent with ultra-small scale quasistatic limit the functions $\zeta$ and $g$ must have the form
$\zeta = \zeta^{(0)} + O(\adotoa_k)$ and $\adotoa_k g =  g^{(-1)} + O(\adotoa_k)$. As discussed in the last part of section \ref{zeta_g_param},
 we further expand $\zeta^{(0)}$ and $g^{(-1)}$ as
\begin{equation}
\zeta^{(0)} = \zeta_1 \Omega_E + \zeta_2 \Omega_E^2 + \zeta_3 \Omega_E^3
\end{equation}
and
\begin{equation}
g^{(-1)} = g_1 \Omega_E + g_2 \Omega_E^2 + g_3 \Omega_E^3
\end{equation}
respectively, where the coefficients may once again be $k$-dependent.
To lowest order in $\adotoa_k$ we  find
\begin{eqnarray}
\Uc_1 &=&   \frac{3}{2} g_1 \nonumber
\\
\Uc_2 &=&   \frac{3}{2} \left[g_2 - g_1 + g_1^2\right] \nonumber
\\
\Uc_3 &=&  \frac{3}{2} \left[ g_3 - g_2 + g_1^2(g_1-1) + 2 g_1 g_2  \right] \nonumber
\\
\Vc_1 &=&  2 g_1 - \zeta_1 \nonumber
\\
\Vc_2 &=&  2g_2+ (2 + 3 w_0) g_1^2  - g_1 \zeta_1 -\zeta_2 \nonumber
\\
\Vc_3 &=& 2 g_3 +2  (1 + 3  w_0) g_1^3 + (3w_1 -3w_0  -  \zeta_1 ) g_1^2 
\nonumber 
\\
&&
+ g_1 g_2 (4 +9w_0) -g_2 \zeta_1 -g_1 \zeta_2 -\zeta_3
\label{UV_param}
\end{eqnarray}
These expressions may then be used with  (\ref{Ytot_param}) and (\ref{gammatot_param})
to find the $\gamma$ coefficients.

\subsection{Comparison with Numerical Results}

The fitting formulas we have derived map theoretical properties of the gravitational field
onto the observable, $\gamma$. This allows us to circumvent the use of a full cosmological perturbation
code when trying to observationally constrain the $\mu^2$, $\zeta$ and $g$ via the growth of
structure. We have already seen how well such an approach fares for the simplified
model we used in Section \ref{small}. We now briefly show how the general framework
fares- note that the model for $g$ and $\zeta$ matches onto the PPN and we have
found the expansion of $\gamma$ to second order in $\Omega_E$. Both of these properties should allow
us to span a wide range of possible models.

\begin{figure}[htbp]
\begin{flushleft}
\vspace{-15pt}
\epsfig{figure=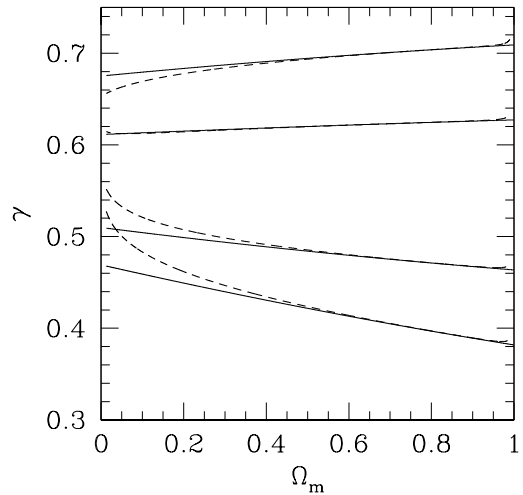,width=9cm}
\end{flushleft}
\vspace{-60pt}
\caption{The growth parameter, $\gamma$, for a selection of dark energy models, as
a function of $\Omega_M$. The dashed curves are the numerical results for
 $\zeta_1=-0.6$, $-0.3$, $0.3$ and $0.6$ in ascending order and the corresponding analytic approximations are plotted in solid line.}
\label{justzeta1}
\vspace{-1pt} 
\end{figure}

\begin{figure}[htbp]
\begin{flushleft}
\vspace{-15pt}
\epsfig{figure=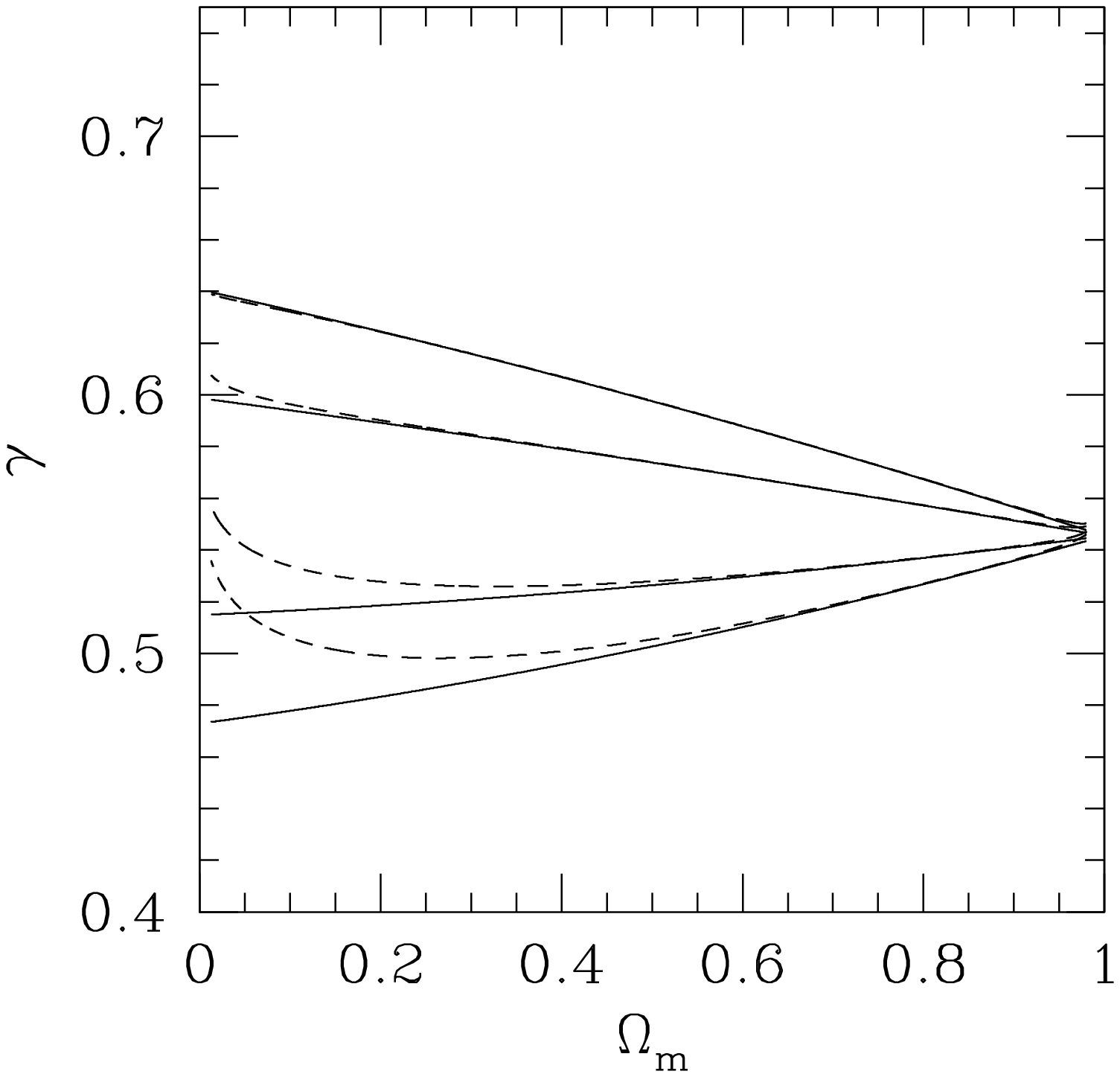,width=9cm}
\end{flushleft}
\vspace{-60pt}
\caption{The growth parameter, $\gamma$, for a selection of dark energy models, as
a function of $\Omega_M$. The dashed curves are the numerical results for
  $\zeta_2=-0.6$, $-0.3$, $0.3$ and $0.6$ in ascending order and the corresponding analytic approximations are plotted in solid line.}
\label{justzeta2}
\vspace{-1pt} 
\end{figure}

\begin{figure}[htbp]
\begin{flushleft}
\vspace{-15pt}
\epsfig{figure=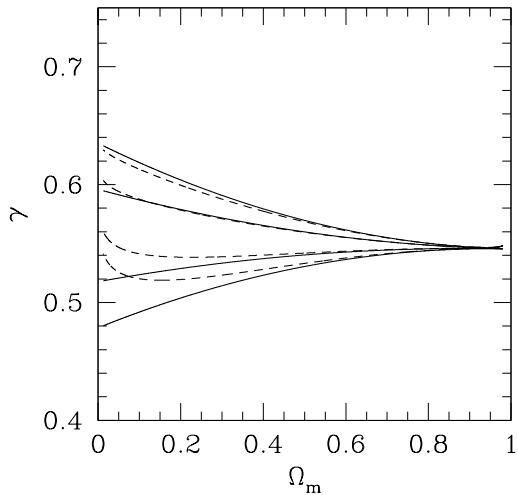,width=9cm}
\end{flushleft}
\vspace{-60pt}
\caption{The growth parameter, $\gamma$, for a selection of dark energy models, as
a function of $\Omega_M$. The dashed curves are the numerical results for
 $\zeta_3=-0.6$, $-0.3$, $0.3$ and $0.6$  in ascending order and the corresponding analytic approximations are plotted in solid line.}
\label{justzeta3}
\vspace{-1pt} 
\end{figure}

If we focus first on the gravitational slip, we can see in Figures \ref{justzeta1}, \ref{justzeta2} and
\ref{justzeta3} that the analytic fit works exceptionally well, to within a few percent out to $\Omega_M = 0.1$. The fit starts to diverge in the upturn of $\gamma$ for low $\Omega_M$ a universal feature that becomes more pronounced for certain ranges of $\zeta$. From inspecting these figures, it is clear that the behaviour of $\gamma$ near $\Omega_E=1-\Omega_M$ is telling
something qualitative about $\zeta$. The parameter $\zeta_1$ drives the offset of $\gamma$ (i.e.$ \gamma_0$),
$\zeta_2$ the slope (i.e. $\gamma_1$) and $\zeta_3$ the curvature (i.e. $\gamma_2$). Measuring
these three coefficents can give a direct handle on the time evolution of $\zeta$.


We find similar effect when looking at the time varying Newton constant, i.e. $g$ (or $\mu^2$).
 Again there is a direct mapping between $g_1$, $g_2$ and $g_3$ and $\gamma_0$, $\gamma_1$ and $\gamma_2$. The accuracy of
the approximation breaks down for smaller values of $\Omega_M$ yet is still excellent in
the range of interest for observational cosmology. For small values of $\Omega_E$ he accuracy is
less than a percent and really only becomes large (of order $5-10\%$) for $\Omega_M<0.1$.

\begin{figure}[htbp]
\begin{flushleft}
\vspace{-15pt}
\epsfig{figure=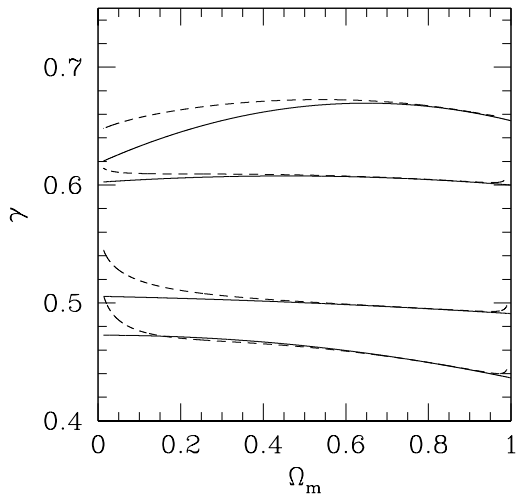,width=9cm}
\end{flushleft}
\vspace{-60pt}
\caption{The growth parameter, $\gamma$, for a selection of dark energy models, as
a function of $\Omega_M$. The dashed curves are the numerical results for
$g_1=-0.4$, $-0.2$, $0.2$ and $0.4$  in descending order and the corresponding analytic approximations are plotted in solid line.}
\label{justg1}
\vspace{-1pt} 
\end{figure}
\begin{figure}[htbp]
\begin{flushleft}
\vspace{-15pt}
\epsfig{figure=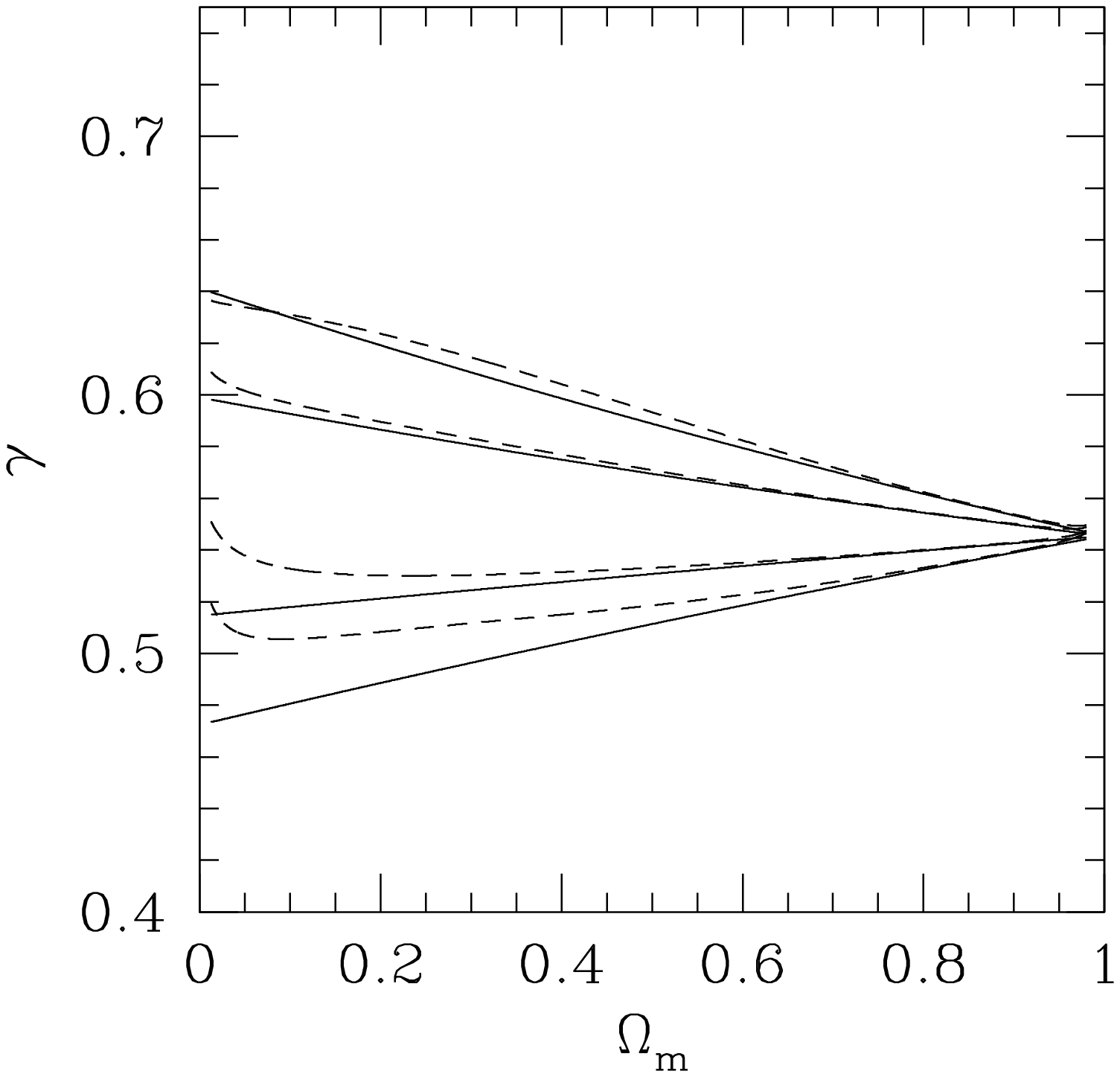,width=9cm}
\end{flushleft}
\vspace{-60pt}
\caption{The growth parameter, $\gamma$, for a selection of dark energy models, as
a function of $\Omega_M$. The dashed curves are the numerical results for
$g_2=-0.4$, $-0.2$, $0.2$ and $0.4$  in descending order and the corresponding analytic approximations are plotted in solid line.}
\label{justg2}
\vspace{-1pt} 
\end{figure}
\begin{figure}[htbp]
\begin{flushleft}
\vspace{-15pt}
\epsfig{figure=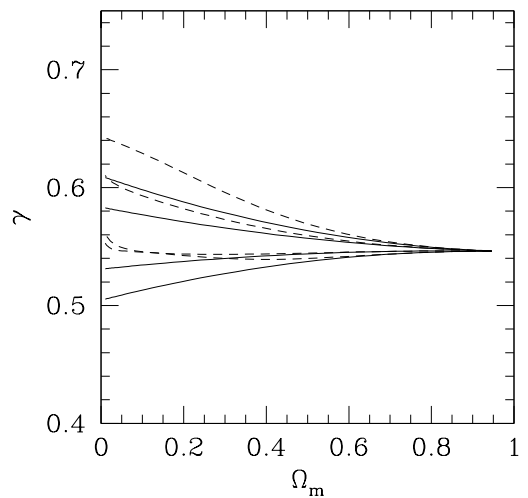,width=9cm}
\end{flushleft}
\vspace{-60pt}
\caption{The growth parameter, $\gamma$, for a selection of dark energy models, as
a function of $\Omega_M$. The dashed curves are the numerical results for
$g_3=-0.4$, $-0.2$, $0.2$ and $0.4$  in descending order and the corresponding analytic approximations are plotted in solid line.}
\label{justg3}
\vspace{-1pt} 
\end{figure}

\section{Discussion}
\label{Discussion}
Let us briefly recap what we have done. The main point of this paper is that, when introducing
modifications to gravity in linear perturbation theory, one must take into account the
consistency conditions in the field equations. These necessarily lead to restrictions in
the form of the modifications that can be introduced. Most notably, and within the context
of second order theories, this means that if one wishes to include modifications to the
Newton-Poisson equation, then one {\it cannot consider the simplified gravitational
slip}, $\Psi=\zeta\Phi$, and must include an extra term such that $\Psi=\zeta\Phi+(g/k){\dot \Phi}$
where $\frac{g}{k}\adotoa=1-G_{eff}/G_0$. If we wish to construct a proper "Parametrized Post Friedmanian" approach to modified gravity, any parameter we introduce must be independent
of the environment or initial conditions in the perturbations. The only way to do this is
to use the parametrization we are advocating. To our knowledge, all attempts at studying
cosmological deviations from general relativity have ignored this and hence it is unclear
what class of theories they map onto and which types of theories  are being constrained.

Having taken this point on board, we have found the expression for the growth
parameter on small scales in terms of both the gravitational slip, $\zeta$ and the modified
Newton constant, $\mu^2=1- \frac{g}{k}  \adotoa$. 
Given a set of cosmological constraints on $\gamma$ and its
dependence on $\Omega_M$, it is now straightforward to calculate constraints on
$\zeta$ and $g$. The growth parameter is given by equation (\ref{gamma_param}) which
can be seen as a Taylor expansion in terms of $ 1-\Omega_M$. The coefficients
in this expansion, $\gamma_0$, $\gamma_1$ and $\gamma_2$ can be expressed
in terms of the equation of state, $w$ (see Equation \ref{w_param}), 
$\zeta$ (see Equation \ref{zeta_param}) and $g$ (see Equation \ref{g_param}) by using 
equations (\ref{UV_param}), followed by equations (\ref{Ytot_param}) and finally equations (\ref{gammatot_param}). 

With these relations in hand, it is now possible to use cosmological observations to place
constraints on theories of modified gravity. In this paper we have focused on small
scales (by which we mean between 1 and 200h$^{-1}$Mpc), scales that should be
probed by redshift space distortion measurements, galaxy power spectra and weak lensing.
Furthermore, we can now do this  consistently, relating modifications in the growth
rate with changes in the gravitational slip. This is of particular importance
when considering weak lensing where observations probe $\Phi+\Psi$. 
It is also clear from our analysis that we have come up against the limitations
of the $\gamma$ parametrization: it is useful and effective on very small scales but not on
scales comparable to  the cosmological horizon. On those scales,
one should be using the full set of field equations. We therefore do not
advocate using our fitting formula to the growth on larger scales such as would be probed
by the Integrated Sach-Wolfe.

How general is this method? We have declared from the outset, the class of theories that we are considering. They
must be metric, with 2$^{\rm nd}$ order equations and satisfy gauge form-invariance. From what
we have learnt about modifications of gravity, these seem a reasonable set of
conditions to apply- they lead to theories which are less likely to be marred
by gross instabilities either at the classical or quantum level. We should point out that
all other attempts at developing such a parametrization have {\it implicitely} made these assumptions
although have not necessarily done so self-consistently.
It is possible to extend this analysis beyond the scope of these theories. 
 If we are to go beyond 2$^{\rm nd}$
order, one must include terms in ${\ddot \Phi}$ or even higher. The Bianchi
conditions will, again, impose a set of constraints on the coefficients of these
terms and should allow a similar type of analysis.

Two well studied theores are worth mentioning. $F(R)$, $F(R^{\mu\nu} R_{\mu\nu})$, etc,  theories come with up-to four time derivatives in
the field equations. Thus they do not fall directly within the methods of this paper but do under the 
general scheme outlined in~\cite{Skordis2008b}. In this case one would have to include terms involving
$\dot{\Phi}$, $\ddot{\Phi}$, $\Psi$ and $\dot{\Psi}$ in to the $G_{00}$ and $G_{0i}$ Einstein equations,
while the $G_{ij}$ equations would need $\dddot{\Phi}$ and $\ddot{\Psi}$ in addition. Theories
with higher derivatives are a subject that warrants further investigation and have yet to be properly
incorporated in any parametrized modifications of standard general relativity.

One other theory, studied extensively is the DGP theory \cite{DvaliGabadadzePorrati2000d}. In this case, only two time derivatives are 
present in the field equations and just like our frame, DGP contains two non-metric dynamical degrees of freedom, which 
can be effectively written ad $\delta_E$  and $\theta_E$. However, our framework cannot encompass DGP because DGP cannot be written as a generalized fluid
as we have assumed of dark energy in this work (hence it is not a failure  of our use of $\delta U_{ab}$).
Nevertheless, our $\gamma$ parameterization in powers of $\Omega_E$ is still valid, and indeed needed.
In the case of DGP we find that 
\begin{equation}
\gamma =    \frac{11}{16}   + \frac{7}{5632} \Omega_E - \frac{93}{4096} \Omega_E^2
\end{equation}
gives an error on $\gamma$ around $5\%$ at $\Omega_M <0.1$, dropping to $2\%$ at $\Omega_M\sim 0.2$ and $<1\%$ for larger values of $\Omega_M$.
The error on the corresponding density contrast at those values of $\Omega_M$ is $<2\%$, $\sim 1\%$ and $<0.5\%$ respectively.
Notice how the coefficients are entirely fixed and {\emph do not} depend on the only free parameter of the theory,
namely the scale $r_c$. Rather $r_c$ comes to play a role only through $\Omega_E = \frac{1}{H r_c}$.


\vspace{-10pt}
\section*{Acknowledgements}
\vspace{-10pt}

We are grateful to Tim Clifton, Tom Zlosnik and Joe Zuntz. This work was 
supported by the BIPAC.

\appendix

\section{The quasistatic limit : connecting to the Parametrized Post Newtonian (PPN) approach}
\label{app_PPN}
We reduce our equations to small scales and slow expansions. To make contact with the PPN expansion,
we will write the Einstein equations in powers of the $3$-velocity $v$.  
The $3$-velocity is related to the $4$-velocity by $u^i = a^{-1} v^i$. In this gauge, for scalar perturbations we get
$u^i =   a^{-1} \grad^i \theta $, so that $v^i =    \grad^i \theta $. Hence  $ \theta = - k^{-2}\grad_i v^i$.
Letting $v = \sqrt{v^i v_i}$ we get on dimensional grounds $ k\theta = v$. 

The PPN order bookkeeping is $ k\frac{\partial}{\partial \tau} \equiv ()' \sim  O(v)$ and $\Phi \sim \Psi\sim  \delta \rho \sim O(v^2)$.
The same bookkeeping prescription holds in our case, and in addition, we also have $\adotoa_k \sim O(v)$ and $\Delta_M \sim \delta_M \sim O(0)$.

We are now ready to find the small scale limit which is consistent with PPN. We start from the operators $A$, $B$, $C_1$ and $C_2$.
Since $\adotoa_k' = - \frac{1}{2} \adotoa_k^2 (1 + 3 w \Omega_E)$ and $\Omega_E' =  - 3 \adotoa_k w \Omega_M \Omega_E$,
and letting $J = \adotoa_k^2 (1 +  w\Omega_E)$ we get
\begin{eqnarray*}
  A &=& -3 \adotoa_k \frac{  2 \adotoa_k (g' - \zeta)  + 2  (   2\adotoa^2_k + \frac{1}{3} ) g  }{1 + \frac{9}{2} J }
\\
   &\rightarrow& 6 \adotoa_k \left[ \adotoa_k \zeta  -  \frac{1}{3}  g  \right]
\end{eqnarray*}

\begin{eqnarray*}
 B &=&   \frac{  2 \adotoa_k (g' - \zeta)  +  \adotoa_k^2  (  1 - 3 w \Omega_E  ) g }{1 +  \frac{9}{2} J }
\\
   &\rightarrow&    \adotoa_k \left[ -2\zeta + 2 g' +  \adotoa_k  (  1 - 3 w \Omega_E  ) g  \right]
\end{eqnarray*}

\begin{eqnarray*}
  C_1
&=&
 - 6 \adotoa_k \frac{ 1 }{1 +  \frac{9}{2} J } \zeta'
+ \frac{1}{(1 +  \frac{9}{2} J)^2 } \bigg\{ 2 + 9 \adotoa_k^2 (1  + 3 w\Omega_E) 
\nonumber 
\\
&&
\ \ \ \
-  27\adotoa_k^4 \left[ 1 +  w\Omega_E  +3 w^2 \Omega_M \Omega_E   \right] 
\bigg\} \zeta  
\\
&&
 + 6 \adotoa_k \frac{1}{1 +  \frac{9}{2} J } g''
 + \frac{3  \adotoa_k^2}{(1 +  \frac{9}{2} J)^2 } \bigg\{4 -6 w \Omega_E 
\nonumber 
\\
&&
\ \ \ \
+ 27 \adotoa_k^2 \left[1+ w \Omega_E  + w^2 \Omega_M \Omega_E \right] \bigg\}g'
\\
&&
 + \frac{3\adotoa_k^3}{(1 +  \frac{9}{2} J)^2} \bigg\{ 1 - 6 w \Omega_E  +9 w^2 \Omega_E
\nonumber 
\\
&&
\ \ \ \
+9   \adotoa_k^2 \left[ 1 - 2 w \Omega_E + 6 w^2  \Omega_E - 9w^2 \Omega_E^2 \right] \bigg\}g
\\
&\rightarrow&   2 \zeta  + 3 \adotoa_k \bigg[ 2g'' +2  \adotoa_k (2 -3 w \Omega_E ) g' 
\nonumber 
\\
&&
\ \ \ \
+ \adotoa_k^2 ( 1 - 6 w \Omega_E  +9 w^2 \Omega_E ) g \bigg]
\end{eqnarray*}
and
\begin{eqnarray*}
 C_2  &=& 3  \frac{  2 \adotoa_k (g' - \zeta)  + 2  (   2\adotoa^2_k + \frac{1}{3} ) g  }{1 + \frac{9}{2} J }
\\
   &\rightarrow& 6  \left[ - \adotoa_k   \zeta  +    \frac{1}{3}  g \right]
\end{eqnarray*}
where $\rightarrow$ denotes taking only the lowest order terms that can contribute to the small scale limit.

Now consider the Einstein equations. In the small scale limit we get
\begin{eqnarray}
&&
-  2 k^2  \Phi - 6 \adotoa (\dot{\Phi} +  \adotoa \Psi)
  = 8\pi G a^2 \rho  \delta 
\nonumber 
\\
&&
\ \ \ \
+  6 \adotoa \left[ \adotoa \zeta  -  \frac{1}{3} k g  \right] \Phi,
\end{eqnarray}
\begin{eqnarray}
&&
 2(\dot{\Phi} +  \adotoa \Psi)
 = 8\pi G a^2 \rho  \theta 
\nonumber 
\\
&&
\ \ \ \
+  \adotoa \left[ -2\zeta + 2 g' +  \adotoa_k  (  1 - 3 w \Omega_E  ) g  \right]  \Phi ,
\end{eqnarray}
\begin{eqnarray}
&&
 6\frac{d}{d\tau}\left(\dot{\Phi} + \adotoa \Psi \right)
+ 12\adotoa (\dot{\Phi} + \adotoa \Psi)
 + 2k^2 (\Phi - \Psi)
=  
\nonumber
\\
&&
\ \ \ \
 -3 (E_F  + E_R) \Psi
+ k^2 \bigg\{  2 \zeta  + 3 \adotoa_k \bigg[ 2g''
\nonumber
\\
&&
\ \ \ \
 + \adotoa_k^2 ( 1 - 6 w \Omega_E  +9 w^2 \Omega_E ) g \bigg]
\nonumber
\\
&&
\ \ \ \
 +2  \adotoa_k (2 -3 w \Omega_E ) g'
   \bigg\} \Phi +  2 \left[ -3\adotoa   \zeta  +      k g \right] \dot{\Phi},
\end{eqnarray}
and
\begin{equation}
\Phi - \Psi =   \zeta \Phi + \frac{g}{k} \dot{\Phi}
\label{red_shear}
\end{equation}

As argued in Section \ref{formalism} we expand $\zeta$ in powers of $\adotoa_k$   and write
\begin{equation}
\zeta = \zeta_L(\Omega_E,k) \adotoa_k^n
\end{equation}
to leading order.
We expand $g$ in a similar way as
\begin{equation}
g = g_L(\Omega_E,k) \adotoa_k^m
\end{equation}
The goal now is to find the smallest powers $m$ and $n$ that can be consistent with the Einstein equations as $\adotoa_k \rightarrow 0$.

It is easily seen that $\zeta' = \zeta_1 \adotoa_k^{n+1}$ for some function $\zeta_1(\Omega_E,k)$ which is found to be 
\begin{equation}
  \zeta_{L1} = -\frac{n}{2}  (1 + 3 w \Omega_E)  \zeta_L - 3  w \Omega_M \Omega_E  \frac{\partial \zeta_L}{\partial \Omega_E}  
\label{zeta_L1}
\end{equation}
Similarly we have $g' = g_{L1}  \adotoa_k^{m+1}$ and $g'' = g_{L2} \adotoa_k^{m+2}$ and similar expressions to (\ref{zeta_L1}) can be found for $g_{L1}$ and $g_{L2}$.

Consider again the Einstein equations and now keep on the lowest orders for each variable. 
For example $\Phi$ is $O(2)$ while $(E_F + E_R)\Psi = O(4)$ and $\ddot{\Phi} = O(4)$ etc.
We get
\begin{equation}
-  2 k^2  \Phi
  = 8\pi G a^2 \rho  \delta 
+  6 k^2 \adotoa_k \left[  \zeta_L \adotoa_k^{n+1}  -  \frac{1}{3}  g_L \adotoa_k^m  \right] \Phi,
\label{red_rho}
\end{equation}
\begin{eqnarray}
 2(\dot{\Phi} +  \adotoa \Psi)
 &=& 8\pi G a^2 \rho  \theta 
+  \adotoa \bigg[ -2\zeta_L  \adotoa_k^n +
 2 g_{L1} \adotoa_k^{m+1} 
\nonumber 
\\
&&
+ (  1 - 3 w \Omega_E  ) g_L \adotoa_k^{m+1}  \bigg]  \Phi,
\label{red_v}
\end{eqnarray}
and
\begin{eqnarray}
\Phi - \Psi
&=&  
 \bigg\{   \zeta_L \adotoa_k^n  
+ 3 \adotoa_k^{m+3} \bigg[  g_{L2}  +(2 -3 w \Omega_E ) g_{L1} 
\nonumber 
\\
&&
\ \ \ \
 + \frac{1}{2} ( 1 - 6 w \Omega_E  +9 w^2 \Omega_E ) g_L  \bigg]
   \bigg\} \Phi 
\nonumber 
\\
&&
+  \left[ -3 \zeta_L \adotoa_k^{n+1}  +       g_L \adotoa_k^m \right] \Phi',
\label{red_P}
\end{eqnarray}
while the shear equation (\ref{red_shear}) remains unchanged.
Clearly the choice $n=0$ and $m=-1$ is consistent with all of the above equations.

Let us investigate whether smaller numbers are possible. 
Suppose that $n<0$. Then the shear equation (\ref{red_shear}) implies
 $ \zeta_L \adotoa_k^n \Phi +   g_L \adotoa_k^m \Phi'=0$, hence if $n<0$ then $m<-1$ and in particular
$m = n -1$. Note that this last relation includes the choice $n=0$ as a special case.
If on the other hand $m<-1$ then the term $g_L \adotoa_k^m \Phi'$ in the shear equation (\ref{red_shear})  is of order less than two which forces
automatically $n<0$. Thus, without loss of generality we may set $m= n-1$. With this choice the Einstein equation (\ref{red_rho})
 becomes
\begin{eqnarray*}
-  2 k^2  \Phi
  &=& 8\pi G a^2 \rho  \delta +  6 k^2 \adotoa_k^n \left[  \zeta_L \adotoa_k^2  -  \frac{1}{3}  g_L  \right] \Phi
\end{eqnarray*}
Therefore if $n<0$ we must have $ \zeta_L \adotoa_k^2  -  \frac{1}{3}  g_L = 0$ and since both $\zeta_L$ and $g_L$ are independent of $\adotoa_k$, this forces
$\zeta_L = g_L = 0$. Thus, the only consistent leading-order choice is $n=0$ and $m=-1$.

To summarize, a consistent small scale limit imposes the expansions
\begin{eqnarray}
\zeta &=& \zeta_L(\Omega_E,k)   + O(\adotoa_k)
\\
g &=& g_L(\Omega_E,k) \frac{1}{\adotoa_k} +  O(\adotoa_k^0) 
\end{eqnarray}

Note that there may be additional constraints on the $k$-dependence of $\zeta_L$ and $g_L$.

One source of worry is the $ \adotoa_k^{-1}$ term that persists in the shear equation (\ref{red_shear}). This is not a problem, however.
We may write the potentials in terms of the matter variables in a way that no ambiguity arises. We find
\begin{eqnarray}
\Phi  &=&  -\frac{4 \pi G a^2 \rho_M}{k^2(1 -g_L) } \delta_M,
\label{eq_Phi_only}
\\
 \Psi&=& 
\frac{1- \zeta_L +  g_L\left( \zeta_L -   g_L + 3  w \Omega_M \Omega_E  \frac{\partial g_L}{\partial \Omega_E} \right)   }{1 - g_L}  \Phi
\nonumber 
\\
&&
-\frac{3}{2}\adotoa \Omega_M \frac{ g_L}{1 - g_L} \theta_M,
\label{eq_Psi_Phi_theta}
\\
 \dot{\Phi} &=& \frac{4\pi G a^2 \rho_M }{1 - g_L} \theta_M 
-  \adotoa  \left[ 1 +  \frac{3  w \Omega_M \Omega_E  \frac{\partial g_L}{\partial \Omega_E}   }{1 - g_L} \right] \Phi,
\end{eqnarray}
which  are perfectly consistent equations.

Finally, (\ref{eq_Psi_Phi_theta}) has a further interesting reduction (in this small scale limit). We replace $\theta_M$ by $-\dot{\delta}/k^2$, then
use $\dot{\delta}_M  = \adotoa f \delta_M$ and finally use (\ref{eq_Phi_only}) to write
$\Phi = \gamma_{PPN} \Psi$
where
\begin{eqnarray}
\frac{1}{\gamma_{PPN}} &=&
 \frac{1}{ 1 - g_L  } \bigg\{
1- \zeta_L 
-  g_L (1 - g_L) f
\nonumber 
\\
&&
+  g_L\left( \zeta_L -   g_L + 3  w \Omega_M \Omega_E  \frac{\partial g_L}{\partial \Omega_E} \right)   
\bigg\}
\end{eqnarray}
while the measured Newton's constant on the Earth, $G_N$ is
\begin{eqnarray}
  G_N    &=&  \frac{G}{\gamma_{PPN}(1 -g_L) }
\end{eqnarray}

Expanding $\zeta_L$ and $g_L$ in powers of $\Omega_E$ we find
\begin{eqnarray}
\gamma_{PPN} &\approx& 1+ \zeta_1 \Omega_E 
\end{eqnarray}
and
\begin{eqnarray}
 \frac{ G_N}{G}    &\approx&  1 +  (g_1 - \zeta_1) \Omega_E 
\end{eqnarray}
hence
\begin{eqnarray}
 \frac{\dot{G}_N}{G_N}    &\approx&    -3w(g_1 - \zeta_1)  \Omega_E   \adotoa
\end{eqnarray}


\bibliographystyle{apsrev}
\bibliography{references}

\end{document}